\def\year{2018}\relax
\documentclass[letterpaper]{article} 
\usepackage{aaai18}  
\usepackage{times}  
\usepackage{helvet}  
\usepackage{courier}  
\usepackage{url}  
\usepackage{graphicx}  
\usepackage[hide]{chato-notes} 
\usepackage{bbm}
\usepackage{amsmath}
\usepackage{pgfplotstable}
\usepackage{threeparttable}
\usepackage{multirow}
\begin{document}

\title{Professional Gender Gaps Across US Cities}
\author{
Karri Haranko\\Aalto University\\karri.haranko@aalto.fi 
\And Emilio Zagheni\\University of Washington\\emilioz@uw.edu
\And Kiran Garimella\\Aalto University\\kiran.garimella@aalto.fi 
\And Ingmar Weber\\Qatar Computing Research Institute\\iweber@hbku.edu.qa}

\maketitle

\begin{abstract}

Gender imbalances in work environments have been a long-standing concern. Identifying the existence of such imbalances is key to designing policies to help overcome them. 
	In this work, we study gender trends in employment across various dimensions in the United States. This is done by analyzing anonymous, aggregate statistics that were extracted from LinkedIn's advertising platform. The data contain the number of male and female LinkedIn users with respect to (i) location, (ii) age, (iii) industry and (iv) certain skills. We studied which of these categories correlate the most with high relative male or female presence on LinkedIn. In addition to examining the summary statistics of the LinkedIn data, we model the gender balance as a function of the different employee features using linear regression. 
Our results suggest that the gender gap varies across all feature types, but the differences are most profound among industries and skills.
A high correlation between gender ratios of people in our LinkedIn data set and data provided by the US Bureau of Labor Statistics serves as external validation for our results.

\end{abstract}

%
%
\section{Introduction}


Gender equality is one of the United Nations Sustainable Development Goals (Goal \#5).\footnote{\url{http://www.un.org/sustainabledevelopment/sustainable-development-goals/}} 
However, achieving a balanced representation of genders in employment remains challenging. 
The World Economic Forum report on Gender gap~\cite{wef2017gendergap}
states that, given current rates of change, it will take 217 years to close the economic gender gap.
To design policies that even out these gender gaps, it is important to identify the areas in which gender gaps are biggest. In this paper, we study the differences between male and female employment across US cities in terms of age, industry and skills. The objective is to discover gender trends across these dimensions and attempt to explain which categories have the greatest effect on the gender gap.

The main source of data for this research is the social networking service LinkedIn. LinkedIn provides an advertising platform which can be used to create and manage advertisements.\footnote{\url{https://www.linkedin.com/ad/accounts}} 
Potential advertisers can specify their desired audience by providing targeting criteria, such as gender, city, industry, or a particular skill. 
Based on these features, LinkedIn provides an estimate of how many LinkedIn users match the criteria. This information can be used to estimate gender balance within any subgroup of professionals by querying and then comparing the audience size for both males and females. 
For example, the platform reports that, globally, the male audience of LinkedIn is 56.4\% of the total combined audience of males and females.\footnote{As of January 2018.\label{x}} 
Similar statistics of employment in the US by gender, age and industry can be found from sources such as the Bureau of Labor Statistics (BLS) and the Census Bureau.\footnote{\url{https://www.bls.gov/cps/tables.htm}} 
However, using digital data from services like LinkedIn has two major advantages: (i) it offers additional information such as skills, and (ii) it is updated continuously as new members join LinkedIn or change jobs, compared to offline sources such as BLS, which are updated once every 5 years.
As an illustrative example, LinkedIn estimates that an advertisement targeted to 18- to 24-year-old males in San Francisco Bay Area with knowledge in Java, has the potential to reach 11,000 people. 
Such estimates of employment and skill data are hard to obtain from any other source.

In this paper, we first validate the anonymous, aggregate LinkedIn data audience estimates by comparing the gender ratios to a dataset from BLS. 
Here we show that there is a strong positive correlation between the two (Pearson's $r$: 0.8, $p$ $<$ $10^{-44}$ ). We then model the gender balance as a function of (i) location, (ii) age, (iii) industry and (iv) skills by using indicator variables and linear regression. The coefficients of the model can be interpreted as the effect that each feature has on the gender balance, relative to a reference category.

%
%
\section{Related Work}





Gender imbalances in access to opportunities such as education, work or career development have 
been a long-standing concern.
International organizations 
have developed indicators for measuring gender equality. 
According to most of these indicators, the progress on gender equality has been relatively slow. According to the World Economic Forum 2017 report on Gender gap, 82 countries have \textit{increased} their overall gender gap compared to last year~\cite{wef2017gendergap}.
Using digital data to understand gender gaps is important in order to inform policy that can mitigate such gaps.
Understanding and quantifying gender gaps has been an active area of research for a few decades~\cite{bimber2000measuring}.
In particular, there has been a lot of effort into understanding gender gaps in employment and hiring practices and the role they play in creating a balanced work environment~\cite{bertrand2001gender,beede2011women}.

LinkedIn has been used before as a source of data in research. Yan et al. (\citeyear{yan2016gender}) used LinkedIn data to analyze gender differences in work duration (number of years worked in a job) in the IT industry and found that
women have shorter work duration. 
Their models predict that the typical `rank and yank'\footnote{\url{http://performance-appraisals.org/faq/rankyank.htm}}
practices in the IT industry help perpetuate this trend.
Zide et al. (\citeyear{zide2014linkedin}) studied LinkedIn profiles in terms of how hiring professionals view them, and how they differ across industries. Based on interviews, they found that women were less likely than men to provide personal information on their profiles.
Altenburger et al. (\citeyear{altenburger2017there}) analyzed 
gender differences in self-promotion choices
on LinkedIn profiles.
They used data from MBA graduates in the US to 
find that women are less likely, relative to men, to utilize data fields that require writing in free-form such as the Summary and Job Description fields. 
On a similar note, Tifferet et al. (\citeyear{tifferet2018self}) studied self presentation differences for 480 LinkedIn profiles from a US city and found that women were more likely than men to signal emotions, whereas men were more likely to signal status.


Related to the use of advertising audience estimates, some studies have been using Facebook as a data source. Chunara et al. (\citeyear{10.1371/journal.pone.0061373}) studied the relationship between obesity prevalence and user interests on Facebook. Gittelman et al. (\citeyear{info:doi/10.2196/jmir.3970}) showed that Facebook likes can be used as predictors of health outcomes and health behaviors to complement traditional public health surveillance systems. Zagheni et al. (\citeyear{zaghenietal17pdr}) showed that Facebook's advertising platform can be used to estimate stocks of migrants.

Compared to the above approaches, our work is one of the first to use LinkedIn data at a large scale to understand gender gap in employment along various dimensions (location, industry, skills, and age).
Our work adds to the existing literature by providing tools that can help collect and analyze recent estimates of gender gaps using LinkedIn data.

%
%
\section{Dataset}

The main data for this research was collected from LinkedIn during January 2018. Their advertising platform\footnote{\url{https://business.linkedin.com/marketing-solutions/ad-targeting}} was used to fetch audience size estimates for different specifications. We queried the service with combinations of five different fields: (i) location, (ii) age, (iii) industry, (iv) skill and (v) gender. Each data instance therefore consisted of a set of five categorical variables, and a number, indicating the estimated size of LinkedIn audience matching those five features.

For this research, the goal is to study the gender gaps in the United States. Thus, we selected 20 US metropolitan areas (\textit{location}) which appeared to have the greatest number of LinkedIn users. 
This enabled us to study gaps at detailed geographical level while avoiding data sparsity. 
Another benefit of studying these US areas was that comparable ground truth data was easily accessible.

LinkedIn's ad platform specifies over 100 industries which are grouped into 17 high-level industry groups. 
We used these 17 top level \textit{industries}. For example, `computer games' and `music' fall into the categories of `media' and `arts', respectively. 
The LinkedIn \textit{skills} section covers hundreds of skills. 
Examples of skills  include `Java programming' and `Military Weapon Systems'. We selected the top 25 hottest skills ranked according to LinkedIn based on the probability of getting hired.\footnote{\url{https://blog.linkedin.com/2016/01/12/the-25-skills-that-can-get-you-hired-in-2016}} 
These skills, mostly in the areas of technology, 
marketing and governance are highly relevant in the modern professional environment. 
Considering these skills can help us understand the impact of such skills on various demographic variables. For example, Frank et al. (\citeyear{frank2017small}) show how advances in artificial intelligence and automation will affect employment in cities.
%
Finally, member \textit{age} was disaggregated into four groups: 18-24, 25-34, 35-54 and above 55. 

In total, we collected audience estimates for all combinations of 20 locations, 4 age groups, 17 industries, 25 skills and 2 genders. 
Additional data was collected with different number of features to allow for some flexibility. For example, we collected male and female counts exclusively for cities. These coarser aggregate data  capture higher level information and bigger audience sizes, while the finer data is more detailed but sparse. All of these different feature combinations were queried, resulting in a final dataset of 98,480 items, out of which 31,782 are non-zero.
%
For our statistical analysis, we excluded data items which had 0 counts for both males and females.

For comparison and validation, we used data from the Bureau of Labor Statistics (BLS). These data contained the percent distributions of employees according to metropolitan area, industry and gender for the year 2014~\cite{bls}. BLS does not provide data at the skill level.

%
%
\section{Measuring Gender Bias}

We begin by defining a few key terms that we use in the rest of the paper.
A \textit{variable} in our case is one of the five categorical features supported by Linkedin -- age, gender, industry, skills and location.
A \textit{population} is a subset of users obtained by using the combinations of various variables, e.g.\ the subset of men aged 18-24 working as software developers in San Francisco. 
Our measure of Gender Balance for a population $p$ is defined as: 

\begin{equation*}
	Gender\ Balance_p = \frac{|male|_p}{|male|_p + |female|_p}.
\label{eq:gender_balance}
\end{equation*}

A value of $Gender\ Balance_p$ of 0.5 indicates a perfect balance between the genders. 
Higher or lower value denotes male or female majority, respectively. 

\section{Modeling Approach}
We start our analysis by validating our dataset.
One of the main issues with using digital data in order to understand societal level trends is that there are a number of biases in such digital traces~\cite{crawford2013hidden,zagheni2015demographic}.
To validate our data, we compared the statistics on locations and industries estimated from LinkedIn to those provided by BLS.\footnote{BLS does not provide data for age groups or skills.} 
From the BLS data, we selected 19 metropolitan areas and 10 industries that have a clear correspondence to the LinkedIn data. 
In cases where the industry names in BLS and LinkedIn do not exactly match, we manually created this mapping,
e.g. `Leisure and hospitality' industry in BLS was mapped to `Recreation, Travel, and Entertainment' in LinkedIn; `Professional and business services' in BLS was mapped to `Corporate Services', etc.
For each metropolitan area, we computed the gender balance for the 10 industries using both the BLS and the LinkedIn datasets.
%

We observed a strong (Pearson) correlation, ranging from 0.63 (Greater San Diego Area) to 0.91 (Greater Chicago Area). 
The overall correlation across all cites combined is 0.80 (p $<$ 0.05 in all cases).
This indicates that, at an aggregate level, our dataset is consistent with traditional surveys. Although this does not guarantee the external validity of our dataset, it indicates that we can rule out large inconsistencies or major skewness. 
Complete results for all the 19 cities are not shown due to space constraints.
Our next step is to test the predictability of gender balance and to evaluate which variables contribute to this. To achieve this goal, we fit a linear regression model 
with the gender balance fraction as the dependent variable $y_i$. 
The other categorical variables (industry, age group, skill and location), are modeled using an indicator variable for each possible value: 

 $y_i = \beta_0 + \beta_1\mathbbm{1}(\text{location$_1$}) + \beta_2\mathbbm{1}(\text{location$_2$}) + \dots 
 + \beta_{21}\mathbbm{1}(\text{age group$_1$}) + \beta_{22}\mathbbm{1}(\text{age group$_2$}) + \dots 
 + \beta_{25}\mathbbm{1}(\text{industry$_1$}) + \beta_{26}\mathbbm{1}(\text{industry$_2$}) + \dots 
 + \beta_{41}\mathbbm{1}(\text{skill$_1$}) + \beta_{42}\mathbbm{1}(\text{skill$_2$}) + \dots + \beta_{66}\mathbbm{1}(\text{skill$_{25}$})
 + \epsilon_i,$
 %

\vspace{2mm}
\noindent 
where $\mathbbm{1}(category)$ denotes a binary indicator variable. 
One category from each of the four groups was left without an indicator variable to avoid 
multicollinearity issues. That is, $k-1$ indicator variables are enough to represent $k$ different categorical variables. These dropped categories act as baselines against which the other categorical variables are compared when interpreting the regression coefficients (shown in Table~\ref{tab:regression}). We selected the category values with the largest total audiences within their groups  as the baselines. These were: (i) 'Greater New York City Area' (location), (ii) '25-34' (age group), (iii) 'Manufacturing' (industry), (iv) 'Corporate Law and Governance' (skill).


%
%
\section{Results}

\begin{table}[]
\centering
	\caption{Results for the linear regression model with the addition of last column showing the Gender balance in each category on LinkedIn. All values except those marked with a + have p-value $<$ $0.0001$. Adjusted $R^2$ was 0.6615.}
\label{tab:regression}
{\resizebox{\linewidth}{!}{
\begin{tabular}{llll}
\hline
\hline
Group                      & Category                                  & Coef. Est.     & Gen. Bal. \\
\hline
\multirow{27}{*}{Skill}    & All skills                                & -0.031          &  \\
                           & Electronic and Electrical Engineering     & 0.416          & 0.882  \\
                           & Virtualization                            & 0.422          & 0.880  \\
                           & Storage Systems and Management            & 0.464          & 0.873  \\
                           & Network and Information Security          & 0.391          & 0.840  \\
                           & Middleware and Integration Software       & 0.434          & 0.837  \\
                           & Algorithm Design                          & 0.479          & 0.833  \\
                           & Shell Scripting Languages                 & 0.403          & 0.825  \\
                           & Mobile Application Development            & 0.475          & 0.816  \\
                           & Perl/Python/Ruby                          & 0.338          & 0.807  \\
                           & Web Architecture and Frameworks           & 0.452          & 0.792  \\
                           & Mac, Linux and Unix Systems               & 0.237          & 0.778  \\
                           & Java                                      & 0.302          & 0.756  \\
                           & Data Engineering and Data Warehousing     & 0.310          & 0.750  \\
                           & Cloud Computing and Distributed Systems   & 0.152          & 0.729  \\
                           & Software Modeling and Process Design      & 0.338          & 0.728  \\
                           & User Interface Design                     & 0.270          & 0.724  \\
                           & Business Intelligence                     & 0.238          & 0.720  \\
                           & Database Management System                & 0.303          & 0.691  \\
                           & Statistical Data Analysis and Data Mining & 0.265          & 0.676  \\
                           & Software QA and Usability Testing         & 0.153          & 0.617  \\
                           & Economics                                 & 0.172          & 0.616  \\
                           & SEO/SEM Marketing                         & 0.062          & 0.606  \\
                           & Corporate Law and Governance              & 0.000          & 0.580  \\
                           & Marketing Management$^+$                  & 0.005          & 0.571  \\
                           & Multi-channel Marketing                   & 0.078          & 0.561  \\
\hline
\multirow{17}{*}{Industry} & All industries                            & -0.095         &        \\
                           & Construction                              & 0.036          & 0.712  \\
                           & Agriculture                               & 0.123          & 0.679  \\
                           & Manufacturing                             & 0.000          & 0.679  \\
                           & High Tech                                 & -0.024          & 0.667  \\
                           & Government$^+$                            & -0.007          & 0.632  \\
                           & Transportation                            & 0.028          & 0.632  \\
                           & Finance                                   & -0.030          & 0.571  \\
                           & Corporate Services                        & -0.054          & 0.564  \\
                           & Consumer Goods                            & -0.029          & 0.556  \\
                           & Media                                     & -0.060          & 0.552  \\
                           & Service Industry                          & -0.033          & 0.548  \\
                           & Organizations and Nonprofit               & -0.046          & 0.529  \\
                           & Recreation, Travel, and Entertainment     & -0.048          & 0.524  \\
                           & Legal$^+$                                 & 0.001          & 0.522  \\
                           & Arts                                      & -0.088          & 0.521  \\
                           & Education                                 & -0.070          & 0.440  \\
                           & Medical and Health Care                   & -0.092          & 0.414  \\
\hline
\multirow{5}{*}{Age}       & All age groups                            & 0.012          &        \\
                           & 55+                                       & 0.205          & 0.667  \\
                           & 35-54                                     & 0.075          & 0.595  \\
                           & 25-34                                     & 0.000          & 0.549  \\
                           & 18-24                                     & 0.045          & 0.531  \\
\hline
\multirow{21}{*}{Location} & All locations                             & -0.037          &        \\
                           & Greater Seattle Area                      & 0.037          & 0.542  \\
                           & Houston, Texas Area                       & 0.059          & 0.542  \\
                           & Greater Los Angeles Area                  & 0.045          & 0.540  \\
                           & Greater San Diego Area                    & 0.083          & 0.540  \\
                           & San Francisco Bay Area$^+$                & -0.013          & 0.537  \\
                           & Dallas/Fort Worth Area                    & 0.056          & 0.533  \\
                           & Greater Denver Area                       & 0.068          & 0.533  \\
                           & Washington D.C. Metro Area$^+$            & 0.005          & 0.533  \\
                           & Austin, Texas Area                        & 0.081          & 0.530  \\
                           & El Paso, Texas Area                       & 0.131          & 0.524  \\
                           & Phoenix, Arizona Area                     & 0.103          & 0.524  \\
                           & Greater New York City Area                & 0.000          & 0.523  \\
                           & Greater Philadelphia Area                 & 0.071          & 0.520  \\
                           & Columbus, Ohio Area                       & 0.117          & 0.517  \\
                           & Greater Boston Area                       & 0.026          & 0.517  \\
                           & Charlotte, North Carolina Area            & 0.098          & 0.515  \\
                           & Greater Chicago Area                      & 0.038          & 0.514  \\
                           & San Antonio, Texas Area                   & 0.131          & 0.513  \\
                           & Indianapolis, Indiana Area                & 0.144          & 0.512  \\
                           & Jacksonville, Florida Area                & 0.109          & 0.508  \\
\hline
                           & Constant                                  & 0.481          &        \\
\hline
\end{tabular}}}
\end{table}


%
%


We first focus our attention on descriptive statistics on gender balance. Column 4 in Table~\ref{tab:regression} shows the gender balance in the LinkedIn data grouped by variable. 
We can observe that:
(i) For most skills, the gender balance is greater than 0.5, meaning, they are male dominant.
This is because, as we mention in the Dataset section, most skills are from the technology industry, which is known to be gender biased.
(ii) Gender balance varies widely across industries. Education and Medical/Health care seem to be the most female dominant categories, whereas Construction and Manufacturing are on the opposite end.
(iii) The proportion of males increases as age increases.
(iv) Metropolitan areas are fairly balanced, as can be noted in the lower part of the table. 

Next, we look at the coefficient estimates in Table~\ref{tab:regression} (column 3). 
These coefficients can be interpreted as the mean change in gender balance when switching from the baseline group to the predictor category, while keeping other indicators constant. 
Negative change denotes greater bias towards women, and positive denotes greater bias towards men. 
For example, we used `Manufacturing' as the baseline category for industries. 
The coefficient of the industry Education is $-0.07$, and can therefore be interpreted that the fraction of men is 0.07 less in Education compared to Manufacturing, on average, with the same locations, age groups and skills as in the baseline.

We notice that many of the coefficients are positive, which suggests that the gender imbalance is highly skewed towards males in the dataset. 
This is likely due to two reasons: 
(i) the chosen skills are strongly biased towards technical industry which means all workers (from other industries) are not equally represented; 
(ii) the skills are also male dominant according to the dataset. 
These two factors lead to many cases where the size of the female audience is 0, but the male audience is around a few hundreds. 
Those cases of maximum gender gap have big impact on the mean of the dataset which shows in the coefficients. 
Most of the negative coefficients can be found among industries, which implies that the chosen baseline (Manufacturing) is one of the most male dominant industries of the group. 
The summary statistics support this view.
%
\note[Kiran]{I am not satisfied with the above paragraph. I am not sure if we should talk about the issues in the data.}
\note[Karri]{I removed the mention about sparsity and moved R2 to table1 caption.}



\note[Kiran]{Should we also say something about the possibility of doing something better with the LinkedIn data (as future work)? like including more skills, and other variables (experience), etc?}

%
%
\section{Conclusion}

Our analysis suggests that the gender gap in employment is fairly similar across locations but varies strongly across industries and, to a lesser extent, across skills. 
Male representation seems to increase with age, possibly indicating generational shifts. A linear regression model supports these findings and offers more details regarding the effect of single variables when holding everything else constant. Compositional changes in the industry structure (and related skills) of US cities may be the driver of differentials in terms gender gaps in employment.  

This study serves as a proof of concept pointing to the untapped value of LinkedIn's advertising platform for social science research. Validation against BLS data suggests that these data are consistent with traditional sources, at an aggregate level. Importantly, LinkedIn data offer new types of information, like prevalence of certain skills, that cannot be easily obtained from traditional surveys. It also offers information about traditional statistics, but available at a finer level of geographic disaggregation, and in a more timely manner, than standard surveys.

\bibliographystyle{aaai}
\bibliography{linkedin-gender-gaps}{} 
\end{document}